# Mechanically Exfoliated Metallic Delafossite $PdCoO_2$ Nanomembranes: Quantum Transport and Electrical Evaluation Toward Interconnect Applications

Chengyu Zhu[1], Pahuni Jain[2], Yi Zhang[2], Junghyun Koo[1], Weideng Sun[1], Yaotian Li[3], Alexander McLeod[3], Chris Leighton[2], and Gang Qiu[1]*

[1]Department of Electrical and Computer Engineering, University of Minnesota, Twin Cities, Minneapolis, MN, 55455, USA

[2]Department of Chemical Engineering and Materials Science, University of Minnesota, Twin Cities, Minneapolis, MN, 55455, USA

[3]School of Physics and Astronomy, University of Minnesota, Twin Cities, Minneapolis, MN, 55455, USA

*Corresponding author. E-mail: gqiu@umn.edu

## Abstract

Metallic delafossite oxides have drawn attention for their ultralow resistivity and coherent electronic transport. However, mesoscopic transport studies are hindered by the limited access to high-quality nanoscale devices. Here, we report single-crystalline $PdCoO_2$ nanomembranes, enabling exploration of quasi-two-dimensional (2D) transport. Shubnikov–de Haas and Aharonov–Bohm oscillations under different magnetic field orientations are observed at low temperatures, from which the electron effective mass and electron phase coherence length are extracted. Beyond quantum transport, the electrical performance of $PdCoO_2$ toward interconnect applications is evaluated. A nearly thickness-independent room-temperature resistivity is observed for flakes down to 40 nm thickness. The nanomembranes exhibit a breakdown current density up to 113 MA $cm^{-2}$, with excellent thermal stability and electromigration resistance. These results demonstrate that mechanically exfoliated $PdCoO_2$ flakes preserve the high crystalline and electronic quality of bulk crystals in the quasi-2D limit, providing a useful platform for mesoscopic transport studies and interconnect applications.

## I. Introduction

The metallic delafossite complex oxides represent a class of highly conductive materials that have attracted substantial interest in recent years due to their exceptional electronic and quantum transport properties[1–8]. Among these materials, $PdCoO_2$ stands out due to its ultralow in-plane resistivity of 2.6 μΩ·cm at room temperature and 8 nΩ·cm at low temperatures, placing it among one of the most conductive materials known to date[6]. Structurally, $PdCoO_2$ features a layered crystal structure consisting of highly conductive triangular Pd layers alternately stacked with insulating Co-O octahedral layers along the *c*-axis, as illustrated in Fig. 1(a). The Pd layers host a metallic electronic structure dominated by delocalized 4*d* electrons, which are primarily responsible for the high in-plane conductivity and metallic transport.

A defining transport characteristic of delafossite oxides is their extraordinarily long electronic mean-free-path. At room temperature, this arises from weak electron–phonon interactions[9]. At low temperatures, the majority of impurities are gettered onto the insulating Co-O layers through a sublattice purification mechanism, leaving the metallic Pd layers ultrapure[10,11] with mean-free-paths up to 20 μm[6]. Under such conditions, electron–electron interactions become increasingly prominent within the 2D Pd layers, leading to a variety of exotic transport phenomena, such as directional ballistic transport[12], electronic hydrodynamics[13], $h/e$ coherent oscillations ($h$ is Planck's constant, and $e$ is the elementary charge)[1], *etc*.

Thus far, transport studies in delafossites have primarily focused on single crystals in bulk form. Mesoscopic delafossite thin films, however, may host rich physical phenomena when their lateral dimensions approach the electron mean-free-path, and their thickness is reduced toward the 2D limit. To this end, several techniques have been used to grow $PdCoO_2$ thin films, such as molecular beam epitaxy (MBE)[14–16], sputtering[17], pulsed laser deposition (PLD)[18,19], and solution-based synthesis[20]. Several of these methods can achieve relatively high crystalline quality, with room-temperature resistivity approaching bulk values after annealing[14,15,17,19,21]. However, the long mean-free-path at low temperatures and the record high residual resistivity ratio (RRR) up to ~440 reported in bulk crystals[10] have not yet been achieved in thin films (where the RRR is <10[15,17,19]), due to defects such as twin boundaries. Thin films to date thus have not been able to access the quantum transport regime at low temperatures. On the other hand, successes have been achieved with mesoscopic transport in micrometer-sized devices defined by focused ion beam (FIB) techniques, using bulk crystals or platelets as starting materials[1,12,13,22–24]. However, the high-energy ion irradiation during the FIB process can degrade the crystalline quality near side walls. Moreover, thicknesses only down to ~5 μm are accessible with this technique[12]. An alternative approach would thus be beneficial to obtain high-quality thinner films for the investigation of quantum transport in the mesoscopic limit.

Compared with the approaches described above, direct mechanical exfoliation from bulk material is a viable alternative that has been widely used in research on 2D van der Waals materials, often yielding large-area, high-quality, thin flakes[25–27]. Although 3D bulk materials are typically difficult to cleave, successful exfoliations have been demonstrated for certain 3D materials, including some oxides (*e.g.*, $Ga_2O_3$)[28]. For $PdCoO_2$, although the interlayer Pd-O bonding is ionic in nature[29], 2D ribbonlike defects exist[30], which may facilitate mechanical exfoliation.

Based on the above, we performed mechanical exfoliation on $PdCoO_2$ single crystals using the standard scotch tape method, yielding high-quality $PdCoO_2$ thin flakes down to 14 nm thickness while retaining excellent crystalline quality. Magneto-transport measurements reveal clear Shubnikov-de Haas (SdH) oscillations under perpendicular magnetic fields, confirming a bulk-like electronic structure. Pronounced Aharonov-Bohm (AB) oscillations are also observed under in-plane magnetic fields, from which we extract exceptionally long phase coherence lengths.

Beyond the fundamental interest in mesoscopic quantum transport, $PdCoO_2$ is also a compelling material platform for next-generation nanoscale interconnects. The continuing demand for higher computing performance has made interconnect scaling increasingly critical, as the resistance in nanoscale wiring has become the major bottleneck for power efficiency and operation speed as transistor dimensions are aggressively reduced. The resistivity of conventional Cu interconnects increases substantially upon scaling, due to enhanced surface and grain boundary scattering, along with the growing fraction of volume occupied by the required diffusion barriers and liners[31]. These challenges motivate the search for alternative conductors capable of maintaining low resistance at reduced dimensions. $PdCoO_2$ is an attractive candidate because of its exceptionally low bulk resistivity, comparable to Cu. In addition, the weak electron-phonon scattering and excellent thermal stability (at least up to 1000 K)[32] of $PdCoO_2$ may help suppress electromigration and improve reliability under high current densities. So far, however, no electrical characterization beyond resistivity has been performed to evaluate the potential of $PdCoO_2$ for interconnect applications. Here, using exfoliated $PdCoO_2$ flakes, the in-plane room-temperature resistivity is confirmed to be around 3.4 μΩ cm for flakes, with no clear thickness-dependence within the thickness range we can electrically access (down to 40 nm). The flakes also have a breakdown current density $J_{BD}$ up to 113 MA cm$^{-2}$ and remain electrically stable under current stressing at 10 MA cm$^{-2}$ for over $10^5$ seconds, suggesting excellent current-carrying capacity and resilience to electromigration. No discernible electrical degradation was observed after annealing at 450 °C, thereby meeting the thermal requirements for back-end-of-line (BEOL) processing. These combined properties establish mechanically exfoliated $PdCoO_2$ as a versatile platform for investigating mesoscopic quantum transport and highlight its potential as a BEOL-compatible, high-current-density interconnect material.

## II. Results and Discussion

The $PdCoO_2$ flakes in this work were cleaved from flux-grown bulk crystals[10,32] (see Supporting Information Section 1 for more details, including growth methods and characterization) using adhesive tapes, before being transferred onto Si/$SiO_2$ substrates with the standard mechanical exfoliation technique[33]. Metathesis/flux-grown crystals were preferred over chemical vapor transport crystals[10] due to their typically thinner initial dimensions. The exfoliated flakes exhibit a quasi-2D morphology, with a typical lateral size of 1 to 10 µm and a thickness of 14 to 300 nm. An atomic force microscopy (AFM) image of a representative 14-nm-thick film and its height profile are shown in Fig. 1(b) and 1(c), confirming a near-atomically-flat 2D surface. Raman spectroscopy further verifies the structural integrity of the exfoliated $PdCoO_2$ flakes (Fig. 1(d)). The characteristic $E_g$ and $A_{1g}$ Raman modes near 520 $cm^{-1}$ and 720 $cm^{-1}$ are detected in both bulk crystal and exfoliated flakes, matching well with the reported peak positions in the literature[18]. A more detailed comparison between bulk and flake Raman spectra reveals that the exfoliated flakes show a relatively enhanced $E_g$ mode intensity with respect to the $A_{1g}$ mode, which can be attributed to an increased contribution from in-plane vibrational modes as the thickness decreases[34]. Taken together, these structural and spectroscopic characterizations confirm that crystallographic integrity is preserved after mechanical exfoliation.

The exfoliated $PdCoO_2$ flakes were patterned into six-terminal Hall-bar devices for resistivity and Hall measurements (see Methods section for more details on device fabrication). An optical image of a representative (40-nm-thick, with lateral size 8 × 1.5 $\mu m^2$) device is shown in the upper-left inset of Fig. 2(a). The longitudinal resistivity $\rho_{xx}$ of this device decreases monotonically with decreasing temperature, from 4.2 µΩ cm at 300 K to 0.32 µΩ cm at 2 K with a corresponding RRR of 13, indicating that the $PdCoO_2$ flakes exhibit metallic transport behavior. We note that while the room-temperature $\rho_{xx}$ values are relatively consistent across different flakes, the low-temperature values vary significantly between flakes (sometimes even reaching negative values, as shown in Supporting Information Section 2). We conjecture that at low temperature, when the electron mean-free-path approaches or exceeds the flake width or length, $\rho_{xx}$ is dictated by ballistic transport and thus becomes sensitive to device-specific geometric factors, including the electrode configuration and flake geometry, making it challenging to accurately determine the low-temperature resistivity, RRR, and mobility discussed below. The magnetic field-dependent transverse resistivity $\rho_{xy}$ as a function of temperature is shown in Fig. 2(b). The low-temperature curves overlap, indicating saturation of the carrier density. The slight non-linearity of $\rho_{xy}$ with respect to the magnetic field at low temperatures may originate from curvature-dependent Hall transport on the nearly hexagonal Fermi surface of $PdCoO_2$[24]. The 2D carrier density in Fig. 2(c) is extracted using $n_{2D} = 1/(eR_H)$ , where $R_H = \mathrm{d}\rho_{xy}/\mathrm{d}B$ is the Hall coefficient averaged over the field range from −9 T to 9 T. At low temperatures, the 2D carrier density is $9.7 \times 10^{16}$ cm$^{-2}$, which converts to an electron density of $1.4 \times 10^{15}$ cm$^{-2}$ per Pd layer after normalizing by the number of Pd layers. This number is very close to the atomic density of the Pd layers ($1.44 \times 10^{15}$ cm$^{-2}$), confirming the monovalent nature of the Pd conduction[35]. The corresponding Hall mobility derived from $\mu = \frac{R_H}{\rho xx(B=0)}$, is shown in Fig. 2(c). It is approximately 500 cm$^2$ V$^{-1}$ s$^{-1}$ at low temperatures in this 40-nm-thick flake and decreases with increasing temperature, as expected due to phonon scattering.

To further understand in-plane electronic transport, magnetoresistance measurements were performed at lower temperatures on another 51-nm-thick Hall bar device, subject to magnetic fields perpendicular to the flake plane, as shown in the inset to Fig. 3(a). In such a configuration, the magnetic field gives rise to SdH oscillations originating from the Pd conduction band. At 300 mK, SdH oscillations are clearly observed around 21 T with a period of roughly 0.016 T, as shown in Fig. 3(a). Fast Fourier transform (FFT) analysis of the SdH oscillations is shown in Fig. 3(b). A single dominant frequency with a peak at 30.2 kT is observed, which matches well with the reported value[13]. The Fermi surface area $A_{\mathrm{F}}$ can be extracted from this oscillation frequency using the Onsager relation $F = \frac{\hbar}{2\pi e} A_{\mathrm{F}}$, where $F$ is the oscillation frequency in the FFT spectrum and $\hbar$ is the reduced Planck constant. The extracted $A_{\mathrm{F}}$ of the $PdCoO_2$ flake is 2.88 Å$^{-2}$ (inset of Fig. 3(b)), in close agreement with previously reported values[12], indicating that the Fermi surface is not significantly modified at the dimensions of this exfoliated flake. The corresponding 2D carrier density $n_{2D}$ was determined from the Fermi surface area using the relation $n_{2D} = \frac{g_s g_v}{(2\pi)^2} A_F$, with spin degeneracy $g_s = 2$ and valley degeneracy $g_v = 1$. The carrier density was determined to be $1.46 \times 10^{15}$ cm$^{-2}$, in very good agreement with the value extracted from Hall measurements.

The oscillation patterns at five different temperatures (from 0.7 K to 3 K) are shown in Fig. 3(c), and the corresponding temperature dependence of the oscillation amplitude is presented in Fig. 3(d). By fitting the experimental data to Lifshitz-Kosevich (LK) theory, *i.e.*, $\Delta R_{xx} \propto \frac{X}{sinhX}$ (where $X = \frac{2\pi^2 k_B m^* T}{e\hbar B}$), a cyclotron effective mass of approximately 1.068 $m_e$ is extracted. This is slightly lower than the previously reported value (1.5 $m_e$)[6], the discrepancy likely arising from fitting inaccuracies due to the low signal-to-noise ratio of the oscillations in these micrometer-sized samples.

Magnetoresistance was further probed in the same device under an in-plane magnetic field parallel to the Pd plane and applied current (Fig. 4(a)). Clear oscillatory patterns are observed between 9 T and 25 T, with a period $\Delta B$ of approximately 1.7 T, as shown in Fig. 4(b). The extracted oscillation period translates to an area of 2437 nm$^2$ following the flux quantization condition $\Delta B = \Phi_0/S$, where $\Phi_0 = h/e$ is the flux quantum. This area matches the product of the sample width $W$ (4 μm in this case) and the spacing $d$ = 5.9 Å between neighboring Pd layers, leading us to attribute the oscillation to the AB effect from an interference loop enclosed by the neighboring Pd layers and the sample boundaries (Fig. 4(a)). This observation is also consistent with previously reported results[1]. When the external magnetic field is tilted away from the sample plane, the oscillation period increases and the amplitude rapidly decreases as the tilt angle increases, eventually vanishing above 10° (Supporting Information Section 3). This can be understood as arising from a small out-of-plane magnetic-field component, which bends the in-plane electron trajectories and eliminates the closed-loop paths required for the self-interference in the AB effect. The temperature dependence of the oscillation pattern is illustrated in Fig. 4(c). The amplitude remains nearly constant below 24 K, then gradually decays with increasing temperature until it eventually vanishes above approximately 50 K.

To quantitatively analyze the temperature dependence of the AB oscillation amplitude, we consider the competition between the mesoscopic sample size and the phase coherence length $L_\Phi$. Oscillation amplitudes were extracted by subtracting a smooth background from the magnetoresistance curves and converting to conductance fluctuations ($\Delta G$). Numerical fitting was then performed in the high-temperature regime (24-50 K) using the relation[36]:

$$\Delta G \propto \exp\left(-\frac{L}{L_\Phi}\right).$$

In this equation, $L$ is the perimeter of the loop ($L = 2W + 2d \approx 8$ μm), and the phase coherence length $L_\Phi$ exhibits an approximate $T^{-1/2}$ dependence in a 2D system[37]. In Fig. 4(d), the oscillation amplitude is plotted as ln ($\Delta G/G_0$) against $T^{1/2}$, where $G_0$ is the zero-field conductivity of the flake at the base temperature. In the high temperature regime ($T > 24$ K), the linear fitting of ln ($\Delta G/G_0$) versus $T^{1/2}$ is shown as the green line in Fig.

4(d). The corresponding $L_\Phi$ in the high temperature regime (Fig. 4(d), right axis), therefore, could be extracted from the fitted slope of ln ($\Delta G/G_0$) versus $T^{1/2}$. At low temperatures ($T$ < 24 K), the oscillation amplitude saturates as $L_\Phi$ approaches the sample size $W$. The crossover near 24 K may reflect a change of dominant transport mechanism in this device. Above this temperature, the lateral electronic transport may gradually evolve from a ballistic or quasi-ballistic regime toward a more diffusive regime, thereby weakening the mesoscopic transport contribution associated with the observed oscillations. Further systematic measurements on devices with different widths and thicknesses are required to verify this interpretation. This model further implies that, in the absence of geometric constraints, the intrinsic coherence length would continue to increase at lower temperatures (e.g., 7.5 μm at 1 K, see Supporting Information Section 4 for more discussions).

In addition to unique quantum transport properties, delafossites have also been considered as promising candidates for next-generation interconnect materials in future microelectronics due to their ultralow bulk resistivity (comparable to the incumbent interconnect material Cu[31]), as well as promising predicted dimensional scaling[38]. However, experimental evaluation of the electrical performance of delafossites has been limited in the context of interconnect applications, particularly in mesoscopic-sized structures.

Exfoliated $PdCoO_2$ nanomembranes were therefore used to evaluate the potential of the material for BEOL interconnect performance metrics, including scalability, electromigration resistance, breakdown current density $J_{BD}$, and thermal stability. The room-temperature resistivity of exfoliated flakes was first measured as a function of thickness. The resistivity was determined using a four-terminal configuration on Hall bar structures. Fig. 5(a) summarizes the room-temperature resistivity extracted from flakes with thicknesses from 40 nm to 100 nm. It should be noted that accurately measuring the resistivity of $PdCoO_2$ is challenging due to the relatively small flake size and ultralow resistance. Although flakes thinner than 40 nm were occasionally obtained by exfoliation (see above), electrical measurements were chalenging due to fabrication or measurement difficulties in such small flakes. To minimize errors arising from geometric nonidealities and fabrication uncertainties, the four-terminal resistivity was determined by averaging the $\rho_{xx}$ values measured from the left and right voltage probe pairs of the six-terminal Hall-bar test structures. The discrepancy between the two sides was typically within 20%, as indicated by the error bars in Fig. 5(a). Detailed measurement data and device geometries are provided in Supporting Information Section 5. Within this thickness range, the resistivity remains stable at 3 to 4 μΩ cm with no discernible thickness dependence. This suggests that surface scattering does not dominate the charge transport in the investigated thickness regime.

Furthermore, current stress measurements were performed under a current density of 10 MA $cm^{-2}$ at room temperature. The resistance remains constant for $10^5$ seconds as shown in Fig. 5(b). In terms of breakdown current density $J_{BD}$ measurements, as the applied current increases, the current-voltage (*I-V*) curve remains linear up to 90 MA $cm^{-2}$, above which a gradual increase in resistance is observed due to Joule heating. Electrical breakdown occurs at a current density of 113 MA $cm^{-2}$ (Fig. 5(c)), a value that significantly exceeds the $J_{BD}$ of commonly used interconnect materials, such as Cu (~ 20 MA $cm^{-2}$)[39], Al (~ 20 MA $cm^{-2}$)[40], and Co (~ 70 MA $cm^{-2}$)[41], highlighting the exceptional current-carrying capability of $PdCoO_2$. In addition to $J_{BD}$, the thermal stability of $PdCoO_2$ was also evaluated under conditions relevant to BEOL processes. Fig. 5(d) shows the *I-V* characteristics measured before and after annealing at 450 °C in Ar, representing the typical thermal requirements for BEOL interconnect processes. The resistance remains nearly unchanged before and after annealing, demonstrating that $PdCoO_2$ retains its electrical integrity under thermal exposure within the BEOL thermal window (~450 °C). Taken together, these features highlight $PdCoO_2$ as a promising candidate for BEOL interconnect applications. One caveat is that these conclusions are based on single-crystalline $PdCoO_2$ flakes, whereas practical $PdCoO_2$ films are likely to be polycrystalline and may exhibit different transport and reliability characteristics due to various forms of disorder, including grain boundaries. Further electrical evaluation of wafer-scale films is therefore required to assess practical integration as an interconnect material.

## III. Conclusions

We have demonstrated mechanically exfoliated $PdCoO_2$ flakes with thicknesses down to 14 nm while maintaining exceptionally high crystalline and electronic quality. The quality of the exfoliated flakes is evidenced by SdH oscillations under out-of-plane magnetic fields and pronounced AB oscillations under in-plane fields, indicating low disorder, long mean free paths, and preserved phase coherence. Beyond quantum transport, $PdCoO_2$ flakes show high potential for interconnect applications. Thickness-independent resistivity is observed down to 40 nm, with an average resistivity of 3.4 μΩ·cm. The high quality of the material is further reflected in the remarkable current-carrying capability of the $PdCoO_2$ flakes, which can sustain current densities up to 113 MA $cm^{-2}$ and remain electrically stable under prolonged current stressing at 10 MA $cm^{-2}$, as well as after thermal exposure up to 450 °C, demonstrating excellent thermal stability. These combined properties highlight mechanically exfoliated $PdCoO_2$ as an attractive candidate for mesoscopic transport studies and BEOL interconnect applications.

## Methods

**Crystal Growth and Flake Exfoliation.** $PdCoO_2$ crystals were synthesized using the metathesis/flux method (Supporting Information Section 1). Small crystals of $PdCoO_2$ were mechanically exfoliated multiple times using an adhesive plastic film from ULTRON SYSTEMS to progressively thin the flakes. The thinned flakes were then transferred onto the target substrate by pressing the tape against the substrate surface.

**Transport and Magnetoresistance Measurements.** Transport and Magnetoresistance measurements were performed using a Quantum Design Dynacool PPMS system with a temperature range of 300 to 1.7 K and a magnetic field range of ±9 T. An AC current of 1 mA with a frequency of 23.33 Hz was injected using a Keithley 6221 current source. The voltage signal was detected with an SR860 lock-in amplifier. Additional high-field magnetotransport measurements were carried out at the Magnetic High Field Laboratory (NHMFL) using a resistive bitter magnet capable of ±31 T measurements. In these measurements, the electrical signal was recorded using SR860 and SR865 lock-in amplifiers.

**Device Fabrication.** Before lithography, the substrates were sequentially cleaned in acetone, methanol, and isopropanol (IPA). Electron-beam lithography was performed using a Raith EBPG 5000 system with ZEP520A as the electron-beam resist. Electrical contacts were defined by electron-beam evaporation of Cr(40 nm)/Au(60 nm), followed by a lift-off process in acetone.

**Electrical Measurements.** Current-voltage (*I-V*) characteristics, stress test, and breakdown test were measured using a probe station in conjunction with a Keithley 2450 source meter. Thermal annealing experiments were conducted on a hot plate in an Ar-filled glove box prior to electrical measurements.

## Figures

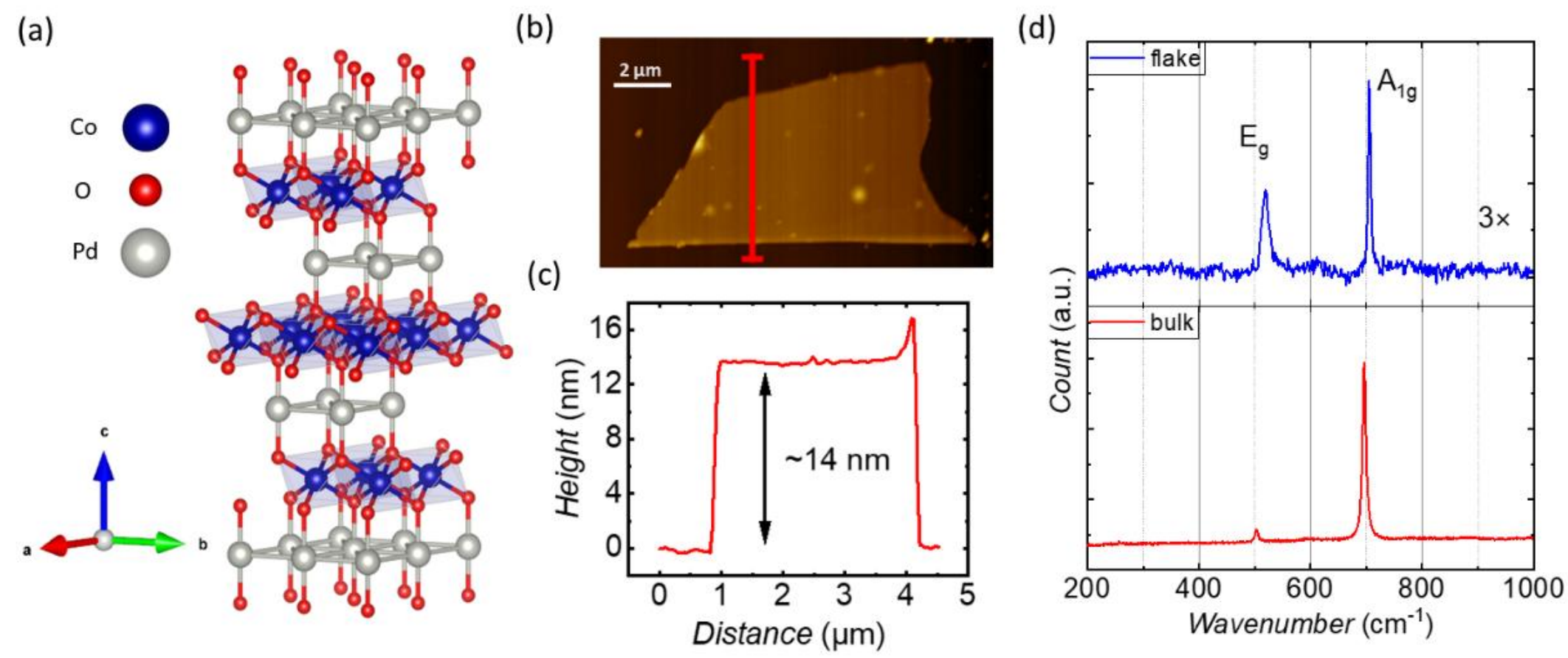

**Figure 1. Structural characterization of exfoliated $PdCoO_2$ thin flakes.** (a) The crystal structure of $PdCoO_2$. Pd, O, and Co are shown in white, red, and blue, respectively. (b) Atomic force microscopy (AFM) image of a mechanically exfoliated $PdCoO_2$ flake. (c) The height profile along the blue line in (b). The flake thickness is 14 nm. (d) Raman spectra of bulk $PdCoO_2$ and a 103-nm-thick flake, showing consistent phonon modes and confirming the structural integrity of the exfoliated flake.

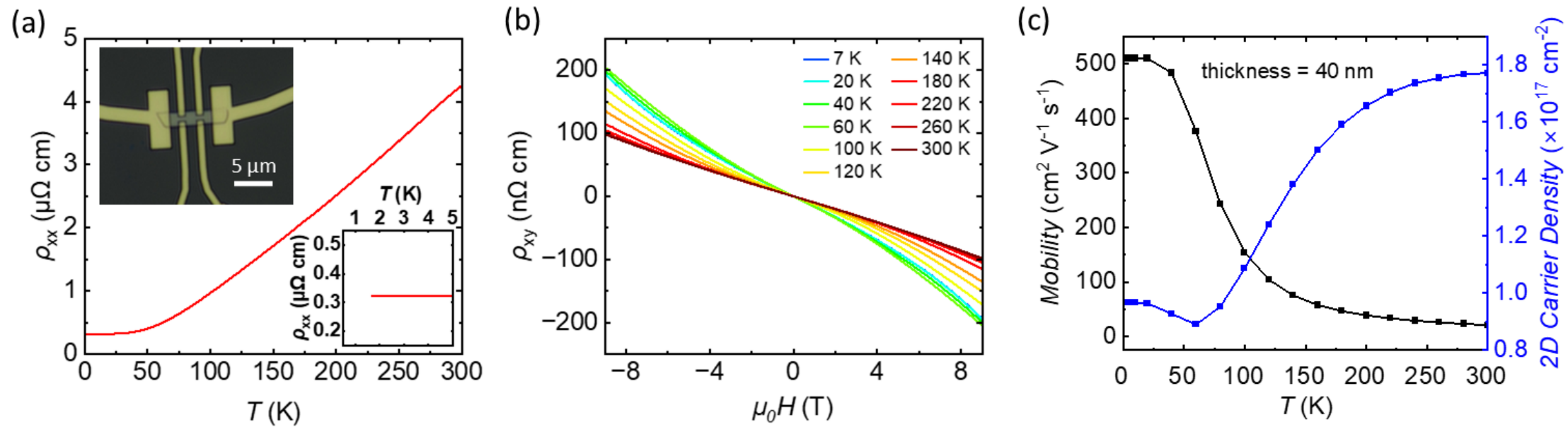

**Figure 2. Hall measurement results for a 40-nm-thick $PdCoO_2$ flake.** (a) Temperature dependence of the longitudinal resistivity $\rho_{xx}$. Inset (upper-left): Optical image of a representative Hall-bar device; (lower-right): zoomed-in $\rho_{xx}$ curve saturating at low temperatures. (b) Transverse resistivity $\rho_{xy}$ as a function of magnetic field $H$ at different temperatures ranging from 7 K to 300 K. (c) Corresponding carrier mobility (black, left axis) and 2D carrier density (blue, right axis) as functions of temperature.

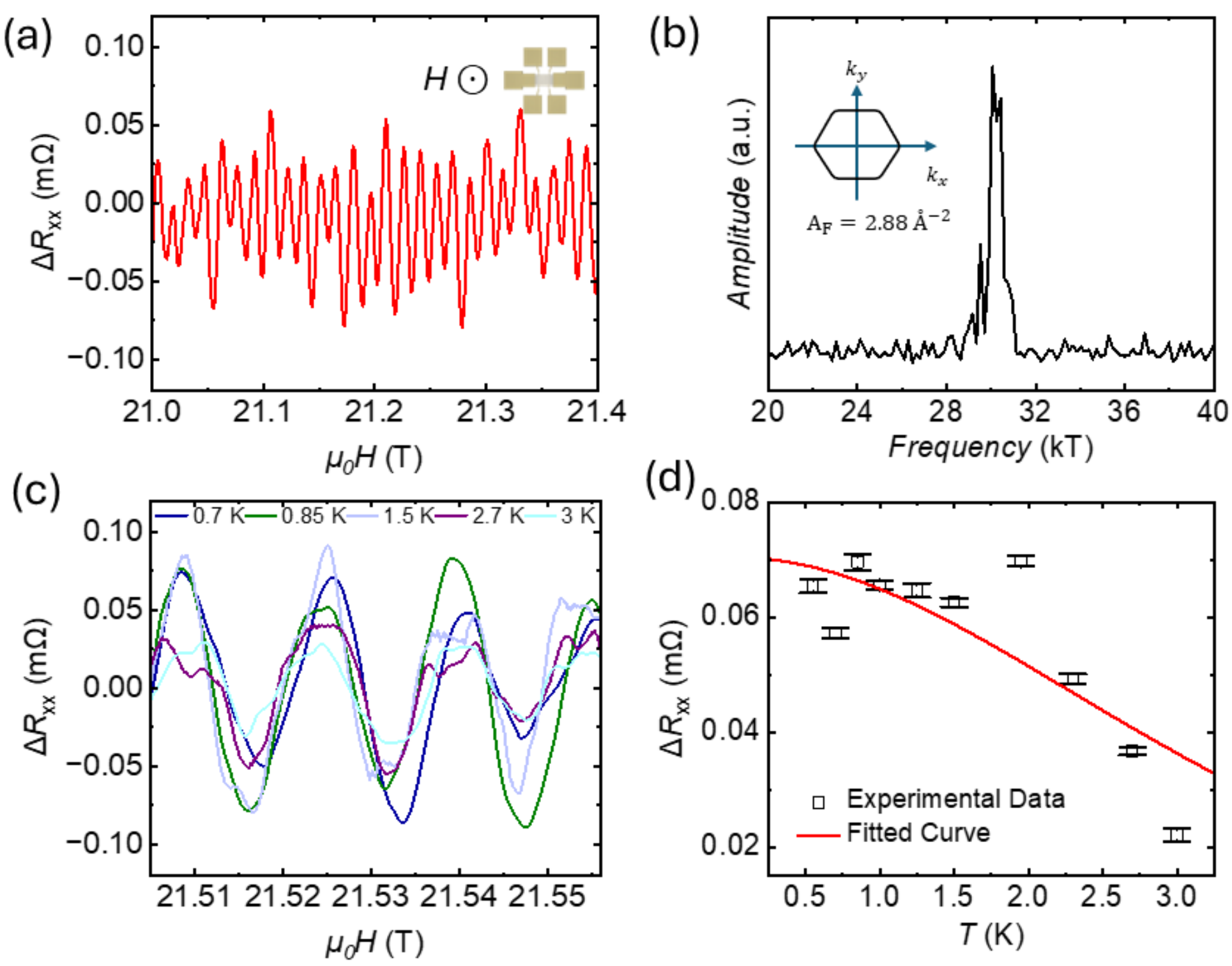

**Figure 3. Shubnikov-de Haas (SdH) oscillations under out-of-plane magnetic field in a 51-nm-thick and 4-µm-wide $PdCoO_2$ flake.** (a) SdH oscillation amplitudes measured in the out-of-plane magnetic field range from 21 to 21.4 T at 0.5 K. A smooth background was subtracted for clarity. (b) The FFT spectrum of the SdH oscillation frequency at 0.7 K, revealing the characteristic oscillation frequency. Inset: the near-hexagonal Fermi surface of $PdCoO_2$ with a surface area of 2.88 Å$^{-2}$. (c) SdH oscillation amplitudes measured between 21.5 and 21.6 T at various temperatures. (d) Temperature dependence of the SdH oscillation amplitude fitted using the L-K formalism. The error bars are from the sinusoidal fitting of the oscillations.

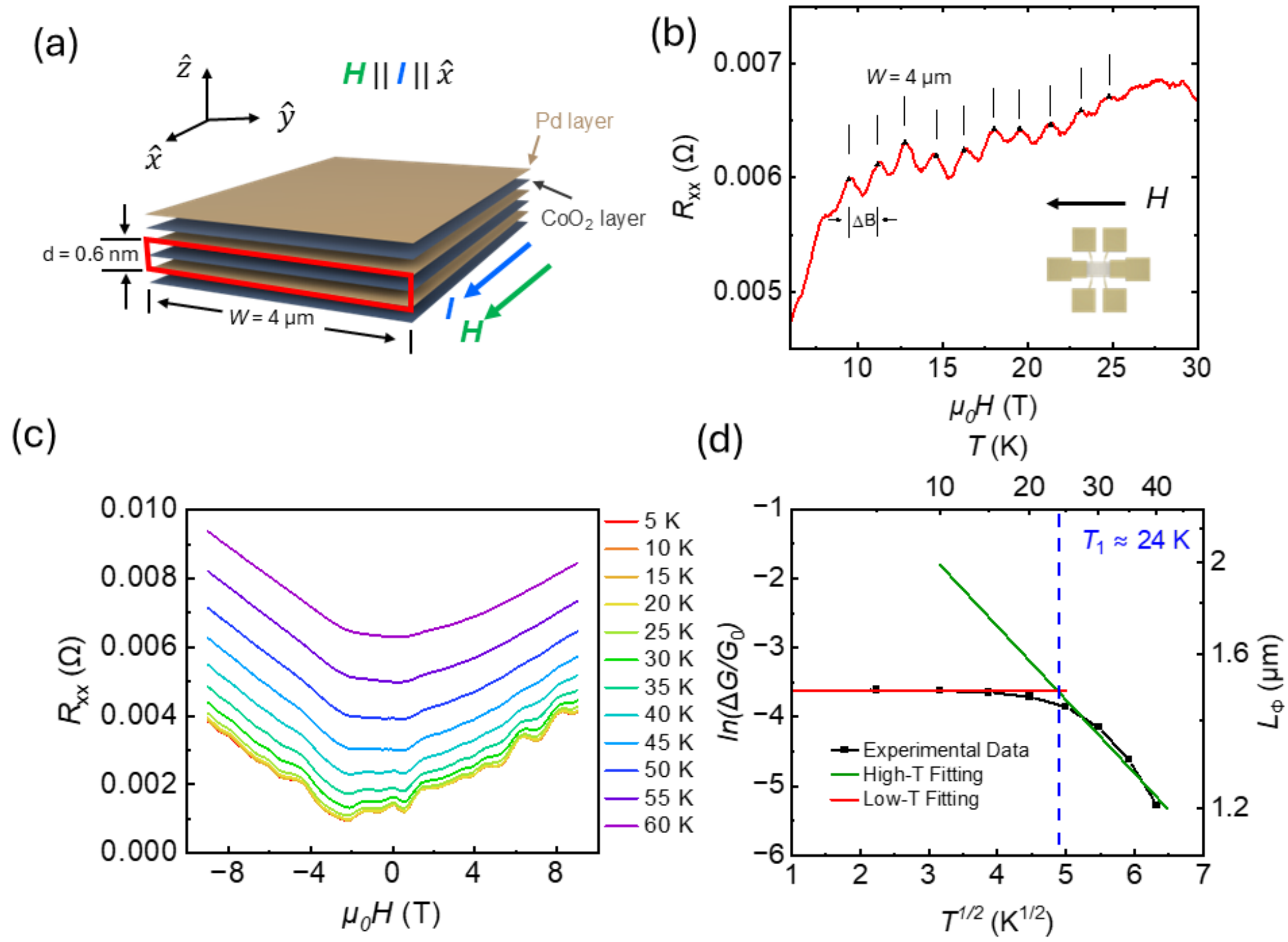

**Figure 4. Aharonov–Bohm (AB) oscillations under in-plane magnetic field in a 51-nm-thick and 4-μm-wide $PdCoO_2$ flake.** (a) Schematic configuration of the experiment. The magnetic field $H$ is parallel to the current $I$. The interference loop for AB oscillations is enclosed by the neighboring Pd layers and the sample boundaries and annotated with the red box. (b) AB oscillation patterns of a flake with a width ($W$) of 4 μm and a thickness ($t$) of 51 nm, ranging from 9 T to 25 T with a period $\Delta B \approx 1.7$ T. Inset: schematic top view of the device geometry and the direction of the in-plane magnetic field. (c) Oscillation patterns at different temperatures. (d) Normalized oscillation amplitude in log scale ln ($\Delta G/G_0$) and Phase coherence length $L_\Phi$ as a function of $T^{-1/2}$. The experimental data (black squares) exhibits two distinct temperature regimes. The green line shows the high-temperature fitting, while the red solid line shows the low-temperature fitting. The blue dashed line ($T_1 \approx 24$ K) indicates the crossover temperature.

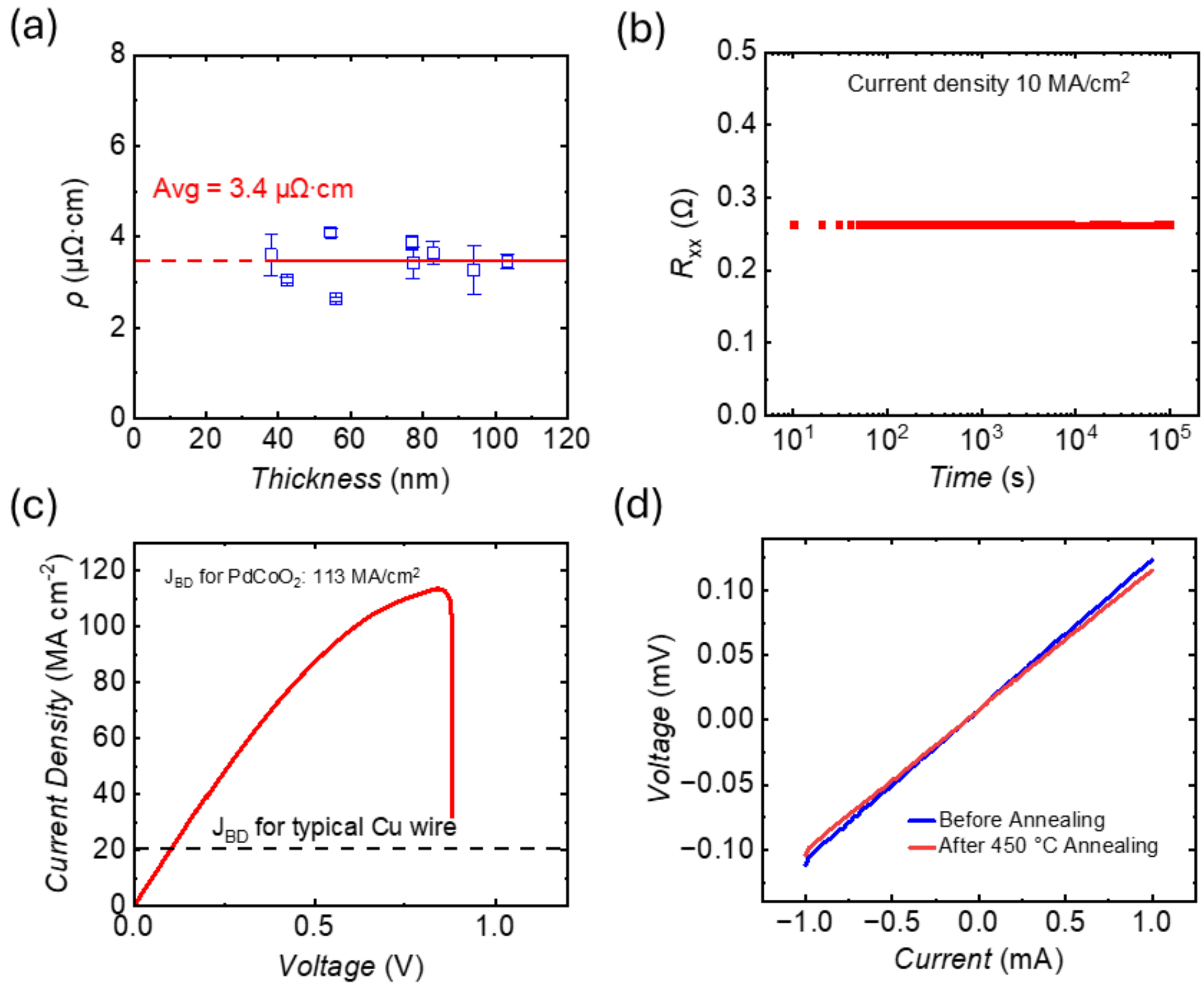

**Figure 5. $PdCoO_2$ Flake Interconnect Evaluation.** (a) Room-temperature resistivity as a function of thickness. The error bars represent the variation in resistance measured from different longitudinal electrode pairs of the same Hall bar device. (b) Resistance as a function of time under a current density of 10 MA $cm^{-2}$ of a 44-nm-thick $PdCoO_2$ flake. (c) Current density *vs*. voltage plot showing the breakdown current density of the same $PdCoO_2$ flake. The dashed line shows the breakdown current of typical Cu wires. (d) Voltage *vs*. current illustrating the resistance of a 77-nm-thick $PdCoO_2$ flake before and after annealing at 450 °C.

## Acknowledgments

G. Qiu acknowledges the Department of Electrical and Computer Engineering and the College of Science and Engineering at the University of Minnesota, Twin Cities for start-up funding support. This research is supported by the Defense Advanced Research Projects Agency (DARPA) through the Young Faculty Award (D25AC00347-00). Work in the Leighton Group (crystal growth and characterization) was also supported by DOE through the UMN Center for Quantum Materials under Grant No. DE-SC0016371. A portion of this work was performed at the National High Magnetic Field Laboratory, which is supported by National Science Foundation Cooperative Agreement No. DMR-2128556 and the State of Florida.

**Conflict of Interest:** The authors have no conflicts to disclose.

## Data Availability Statement

The data that support the findings of this study are available from the corresponding author upon reasonable request.